# Inverse designed metalenses with extended depth of focus

*Elyas Bayati[1,\*], Raphael Pestourie[2,3,\*], Shane Colburn[1], Zin Lin[2], Steven G. Johnson[4], Arka Majumdar[1,4,+]*

[1] *Electrical and Computer Engineering, University of Washington, Seattle, WA 98189, USA*
[2] *Department of Mathematics, MIT, Cambridge, MA 02139, USA*
[3] *Harvard John A. Paulson School of Engineering and Applied Sciences, Harvard University, Cambridge, MA 02138, USA*
[4] *Department of Physics, University of Washington, Seattle, WA 98189, USA*
[\*] *These authors contributed equally*
[+] *Corresponding Author: arka@uw.edu*

**Abstract**

Extended depth of focus (EDOF) lenses are important for various applications in computational imaging and microscopy. In addition to enabling novel functionalities, EDOF lenses can alleviate the need for stringent alignment requirements for imaging systems. Existing designs, however, are often inefficient or produce an asymmetric point spread function (PSF) that blurs images. Inverse design of nanophotonics, including metasurfaces, has generated strong interest in recent years owing to its potential for generating exotic and innovative optical elements, which are generally difficult to design based on intuition alone. Using adjoint optimization-based inverse electromagnetic design, in this paper, we designed a cylindrical metasurface lens operating at ~

625 nm with a depth of focus exceeding that of an ordinary lens. We validated our design by nanofabrication and optical characterization of silicon nitride metasurface lenses (with lateral dimension of 66.66 μm) with three different focal lengths (66.66 μm, 100 μm, 133.33 μm). The focusing efficiencies of the fabricated extended depth of focus metasurface lenses are similar to those of traditional metalenses.

**TOC Figure:**

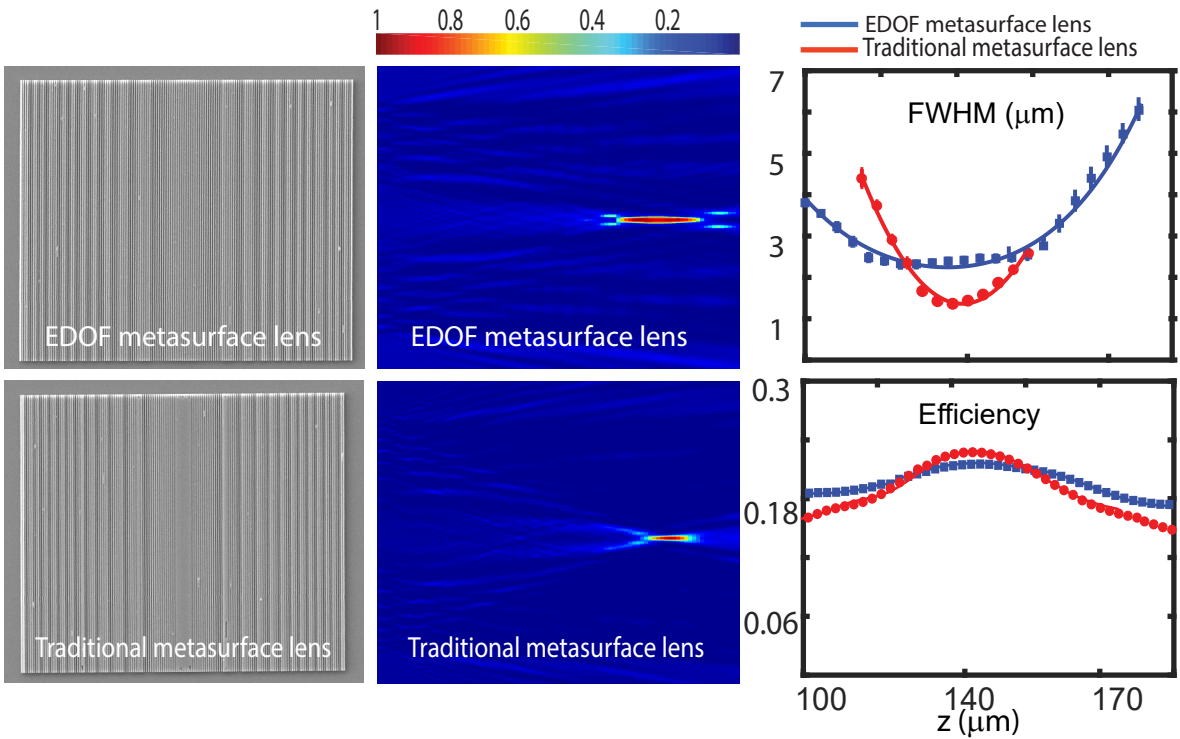

**Main Text**

Sub-wavelength diffractive optics, also known as metasurfaces, have generated strong interest in recent years due to their ultrathin nature and extensive capabilities for manipulating optical wavefronts[1,2]. By virtue of a large number of scatterers, each capable of shaping the phase, amplitude, or polarization of light, metasurfaces can create ultra-compact optical elements, including lenses[3-5], polarization optics[6], axicons[7], holograms[8,9] and freeform optics[10]. While metasurfaces provide an extremely large number of degrees of freedom to design complex optical functions, our intuition often fails to harness all these degrees of freedom. One promising solution is to employ computational techniques to design metasurfaces, where the design process starts from the desired functionality, and the scatterers are designed based on a specified figure of merit. Such design methodologies, often referred to as inverse design, have been employed to design high efficiency periodic gratings[11, 12], monochromatic lenses[13, 14], PSF-engineered optics[15] and achromatic lenses[16-18]. The inverse design is important because it could outperform the conventional approach for the design of metasurfaces, specifically, when desired function is not possible to design with conventional method. Till date, however, there are limited experimental demonstrations of inverse-designed aperiodic metasurfaces exhibiting superior performance to optics designed by traditional methods. For example, recently demonstrated inverse-designed cylindrical lenses exhibit high efficiency[19], though the efficiency is not higher than that of a traditional refractive lens.

Extended depth of focus (EDOF) lenses represent an important class of optical elements, with significant utility in microscopy[20] and computational imaging[21]. This class of lenses differs from an ordinary lens, as the point spread function (PSF) of the lens remains the same over an extended distance along the optical axis. EDOF lenses not only enable novel functionalities, such as bringing objects at different distances away from the lens into focus but also alleviate the stringent requirements on aligning lenses on top of a sensor. While for an ordinary lens the gap between the lens and the sensor plane needs to be very close to the focal length, with EDOF lenses, the gap can vary to a degree without sacrificing the performance of the imaging system. In recent years, metasurfaces have been employed to implement two-dimensional EDOF lenses[22, 23]. Existing EDOF lenses, however, have several problems. One of the most prevalent classes of EDOF lenses is created via a cubic phase mask positioned at the exit pupil[20]. Such an approach generates an Airy beam, which propagates through free space without significant distortion. The resulting PSF, however, does not resemble a point, and images captured with this element are therefore blurry. Computational reconstruction is required to undo the distortion. Another option could be to use a log-asphere lens or other variants[24]. For these lenses, however, different parts of the of the lens focus at different depths, significantly limiting the focusing efficiency. In one dimension, several works theoretically explored cylindrical EDOF lenses [25-28], although to the best of our knowledge no experimental demonstration has previously been reported.

In this paper, we design and fabricate an EDOF cylindrical metasurface lens (metalens) using an inverse electromagnetic design methodology. Unlike existing implementations of diffractive or refractive EDOF lenses, the reported metalens creates a lens-like PSF, without introducing significant blur like many other EDOF lenses. We designed the EDOF lens to have three times the depth of focus of an ordinary metalens. The design is experimentally validated by fabrication and optical characterization. We found reasonable agreement between the simulated and experimental results in terms of the focusing efficiency and the FWHM of the focal spots. The depth of focus of the inverse designed metalens is extended by a factor of $\sim 1.5 - 2$ over that of the traditional metalens. We also did not observe any degradation of the efficiency in the designed EDOF lenses.

**Inverse design and validation**

The depth of focus $\Delta f$ of an ordinary lens with diameter $D$ and focal length $f$, for optical wavelength $\lambda$ is given by

$$\Delta f = 4\lambda \frac{f^2}{D^2}$$

We aim to demonstrate an EDOF metalens with a depth of focus of three times this value. We specify our figure of merit (FOM) as the intensity at eight linearly spaced points along the optical axis which cover an interval of length $3 \times \Delta f$ and that are centered around the focal length. We use max-min multi-objective optimization[29] to maximize the intensity uniformly on the segment. We optimize the EDOF lens FOM by adjusting the widths of 150 nano-stripes positioned at the center of each lattice cell. The material of the metasurface is assumed to be silicon nitride (n~2). The lattice periodicity and thickness of the metasurface are kept constant at 443 nm and 600 nm, respectively. The choice of these parameters is dictated by the ease of fabrication, such as the total area of fabrication and reproducibility of the feature sizes. To ensure the fabricability of the designed metasurface, we constrain the minimum width of the stripes to 100nm. Lenses with three different focal lengths (66.66 µm, 100 µm, 133.33 µm) are designed. To channel most of the light into the main lobe, i.e., to reduce the sidebands in the focal length we ran the inverse design with different initial conditions, which resulted in several different geometries. We then chose the design with the least power in the sidebands. In contrast to a traditional lens, the inverse problem of designing an EDOF lens has multiple solutions to Maxwell's equations that could be good candidates for the EDOF design. Our inverse design framework enables us to set the problem without constraining ourselves to just one of these solutions. Note that if our design method was based on optimizing a phase mask instead of solving for the geometry of our scatterers directly, we would be limited to a single solution, which would be suboptimal in the sense that it limits the degrees of freedom. We note that, however, an inverse design method, where the refractive index of each voxel in the structure are used as degrees of freedom can potentially provide a better solution, albeit with significantly increased computational time. As such, the inverse design method employed here provides a good trade-off between an improved design and computational complexity.

**Experimental Demonstration**

To validate our metasurface design, we fabricated the cylindrical metalenses in silicon nitride. A 600-nm-thick layer of silicon nitride was first deposited on a 500-um-thick fused-silica substrate using plasma-enhanced chemical vapor deposition (PECVD). The sample was then spin-coated with electron-beam resist (ZEP-520A) and then the metalens' patterns were exposed via electron-beam lithography. 8-nm of Au/Pd as a charge dissipation layer was sputtered on the resist prior to exposure to prevent pattern distortion due to electrostatic charging. After the lithography step, the charge dissipating layer was removed by type TFA gold etchant and the resist was developed in amyl acetate. A 50 nm layer of aluminum was then evaporated onto the sample. After performing lift-off, the sample was etched using an inductively coupled plasma etcher with a mixture of $CHF_3$ and $O_2$ gases, and the remaining aluminum was removed in AD-10 photoresist developer. To demonstrate the extension of the depth of focus, we fabricated two sets of lenses: one set of ordinary metalenses designed via the forward design method[10] and another set of EDOF metalenses developed using our inverse design method[29]. The forward design method of a metalens involves selecting the appropriate spatial phase profile for the specific optical component, arranging the scatterers on a subwavelength lattice, and spatially varying their dimensions. Whereas, in the inverse design, no a priori knowledge of the phase distribution is assumed, and the metalens is designed via optimizing the FOM. Figs. 1A, B show the scanning electron micrograph (SEM) of the fabricated EDOF metalens (using inverse design method) and traditional metalens (using forward design method), respectively. Figs. 1C, D show a zoomed-in SEM of the inverse-designed EDOF metalens, which shows silicon nitride nano-stripes forming the cylindrical EDOF metalens. We fabricated three metalenses corresponding to three different focal lengths, in each set. The fabricated lenses were measured using a confocal microscopy setup under illumination by a 625 nm light-emitting diode (part number Thorlabs-M625F2), see Fig. 1E. Figs. 2A-F show the simulated and experimentally measured field profiles for the three inverse-designed EDOF metalenses. The field profiles are simulated using 2D finite-difference time-domain (FDTD) simulation with an axial sampling resolution of 50 nm. The intensity profiles along the optical axis are captured using a camera and translating the microscope along the optical axis using an automated translation stage with an axial resolution of 2 µm. We find that the simulated field profiles from the designed structures match quite well with the experimentally measured focusing behavior. A clear elongation of the focal spot along the optical axis is observed. We also characterized the performance of inverse-designed EDOF metalens under oblique incidence angles (5°, 10° and 15°) using FDTD simulation (see supplementary materials and Figure S2). While we clearly observe the effect of off-axis aberrations, the extended depth of focus remains the same for different angles. We emphasize, however, that as the inverse design figure of merit did not explicitly handle nonzero incident angles, we do not expect such aberrations to be mitigated in our design. We fit the intensities near the focal plane using a Gaussian function to estimate the full-width-half-maxima (FWHM). Figs. 2G-O show the Gaussian fit focal spot at the center focal plane, and at two ends of the line along which the beam profile starts to become double Gaussian. We identified the depth of focus as the range along the optical axis, where the beam profile remains Gaussian. The minimum FWHM for fabricated EDOF metalenses with three different focal lengths (66.66 µm, 100 µm, 133.33 µm) which are shown in Figs. 2H, K, N are 1.07 µm, 1.7 µm, 2.32 µm, respectively. In Fig. 3 we plot the cross-sections of the focal plane of the EDOF and traditional metalens to compare the PSFs. Clearly the PSF for both lenses look similar, although the FWHM is slightly larger for the EDOF metalenses.

By plotting the FWHM as a function of the distance along the optical axis, we estimated the focal length of the metalenses (Figs. 4A-C). We then estimated the focusing efficiency of the lenses along the optical axes. We define the focusing efficiency as the power within a circle with a radius of three times the FWHM at the focal plane to the total power incident upon the metalens[3, 5]. The FWHM at the focal plane is calculated as the minimum FWHM from the Figs. 4A-C. We plot the focusing efficiency of the metalenses along the optical axis (Figs. 4D-F). We expect the focusing efficiency to remain the same along the depth of the focus, and then drop off as we longitudinally move away from the depth of focus. Clearly, for the EDOF metalens, the efficiency remains high over a longer depth as expected. Table-I summarizes the performance of all the metalenses, in terms of FWHM, focal length, efficiency, and depth of focus. We find a reasonable agreement between the simulation and experimental results. Additionally, we clearly observed an extended depth of focus in the inverse-designed metasurfaces compared to the ordinary metalenses. We note that, in simulation we sample along the optical axes more finely compared to the experiment, which determines a larger error bar in experimentally measured depth of focus. We also observe no significant efficiency degradation between the ordinary metalenses and the EDOF metalenses. Other potential EDOF alternatives (e.g., axicons, log-aspheres, or cubic functions) also exist for 1D lenses; however, different regions of these lenses focus at different depths, significantly limiting the focusing efficiency. We designed and simulated these EDOF cylindrical lens metasurface alternatives to compare their performance with our inversed designed structure. To make a fair comparison with one of our lenses, other 1D EDOF lenses are designed to have central focal spot at 100 μm and a depth of focus of 30 μm (see the supplementary materials and Figure S1 for the simulated field intensity along the optical axis of these elements and corresponding PSFs). The focusing efficiencies of 1D log-asphere, axicon, and cubic lens were 12.16%, 8.6% and 14.47%, respectively. These efficiencies are significantly lower than that of our inverse designed EDOF lens.

**Discussion**

We demonstrated inverse designed EDOF cylindrical metalenses for the first-time. While several theoretical designs exist for cylindrical EDOF lenses[25-28], to the best of our knowledge no experimentally demonstrated EDOF cylindrical lenses have been reported before. While we extended the depth by a factor of $\sim 1.5 - 2$, the depth of focus can be further extended albeit at the cost of reduced efficiency. In this work, we focused on 1D cylindrical lenses for the simplicity of design, cylindrical lenses are similar to spherical lenses in the sense that they converge or diverge light, but they have optical power in only one dimension and will not affect light in the perpendicular dimension. Consequently, they are not suitable for 2D imaging. Additionally, there are many other relevant factors for high-quality imaging including geometric aberrations, getting commensurate PSFs for different incident angles and good efficiency which are not specified in our figure of merit. However, we believe that the demonstrated inverse design method can incorporate a more complicated figure of merit to create a metasurface more suitable for imaging. Thus, the next step will be to extend our inverse design concept to 2D lenses and demonstrate imaging[30-33] over a broader optical bandwidth than what is possible using a traditional metalens using a more complex figure of merit. Going beyond an extended depth of focus, the inverse design techniques can be used to engineer the PSF for other functionalities with potentially far-reaching impacts in computational imaging and microscopy.

**Funding Sources**

This work is partially supported by Samsung-GRO, UW reality lab, Google, Huawei and Facebook, NSF-1825308, U. S. Army Research Office through the Institute for Soldier Nanotechnologies (W911NF-13-D-0001). Part of this work was conducted at the Washington Nanofabrication Facility / Molecular Analysis Facility, a National Nanotechnology Coordinated Infrastructure (NNCI) site at the University of Washington, which is supported in part by funds from the National Science Foundation (awards NNCI-1542101, 1337840 and 0335765), the National Institutes of Health, the Molecular Engineering & Sciences Institute, the Clean Energy Institute, the Washington Research Foundation, the M. J. Murdock Charitable Trust, Altatech, ClassOne Technology, GCE Market, Google and SPTS.

**Supporting Information Available:**

Comparison with other potential EDOF lenses

Simulated performance under oblique incidence

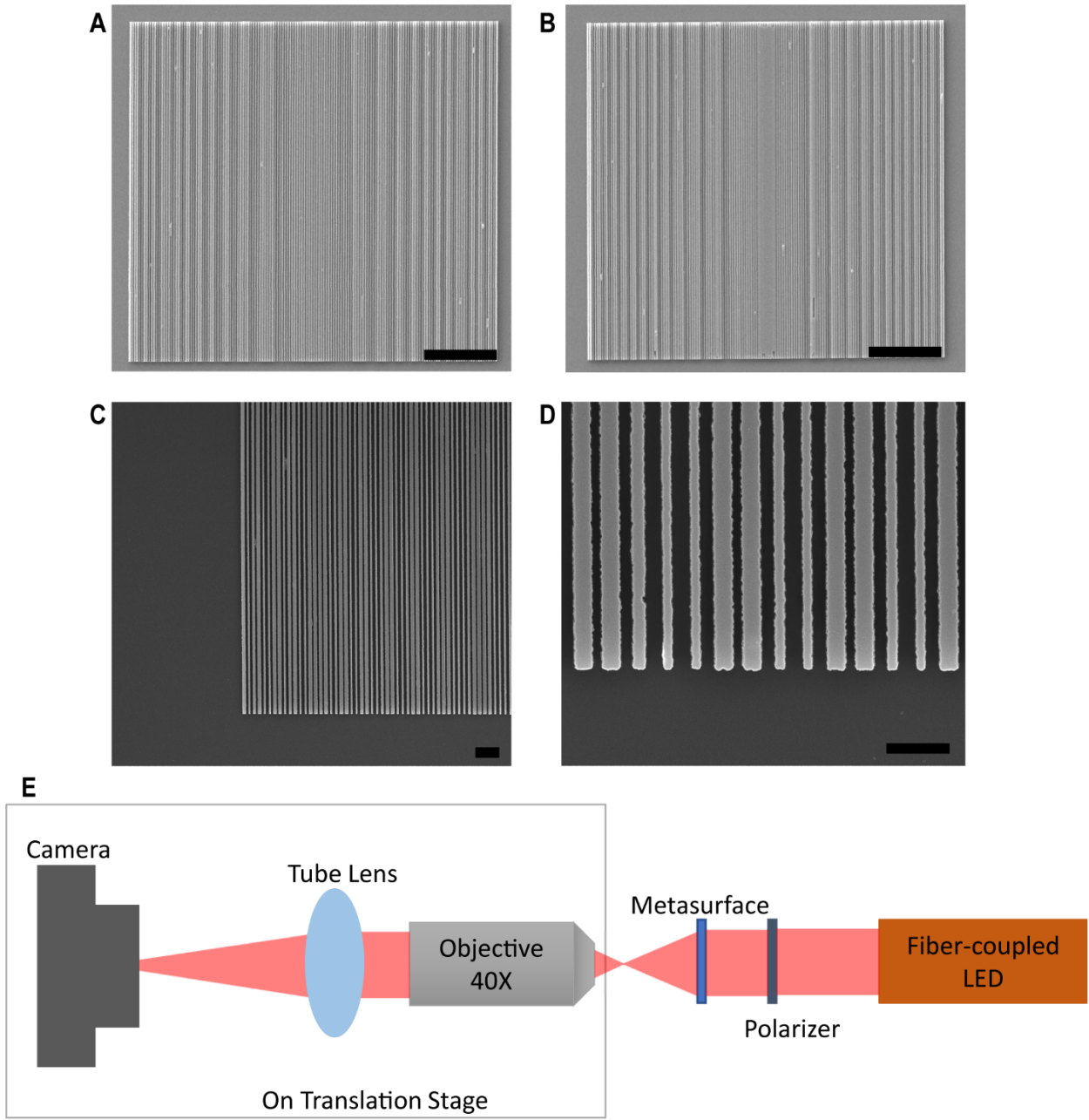

**Figure 1**: Scanning electron micrograph (SEM) of (A) inverse-designed EDOF metalens, (B) traditional metalens; (C) and (D) Zoom-in SEM on inverse-designed EDOF metalens which shows silicon nitride stripes forming the cylindrical EDOF metalens; The scale bars correspond to 10 μm (A-B) and 1 μm (C-D), (E) Confocal microscopy setup used to measure the metalenses.

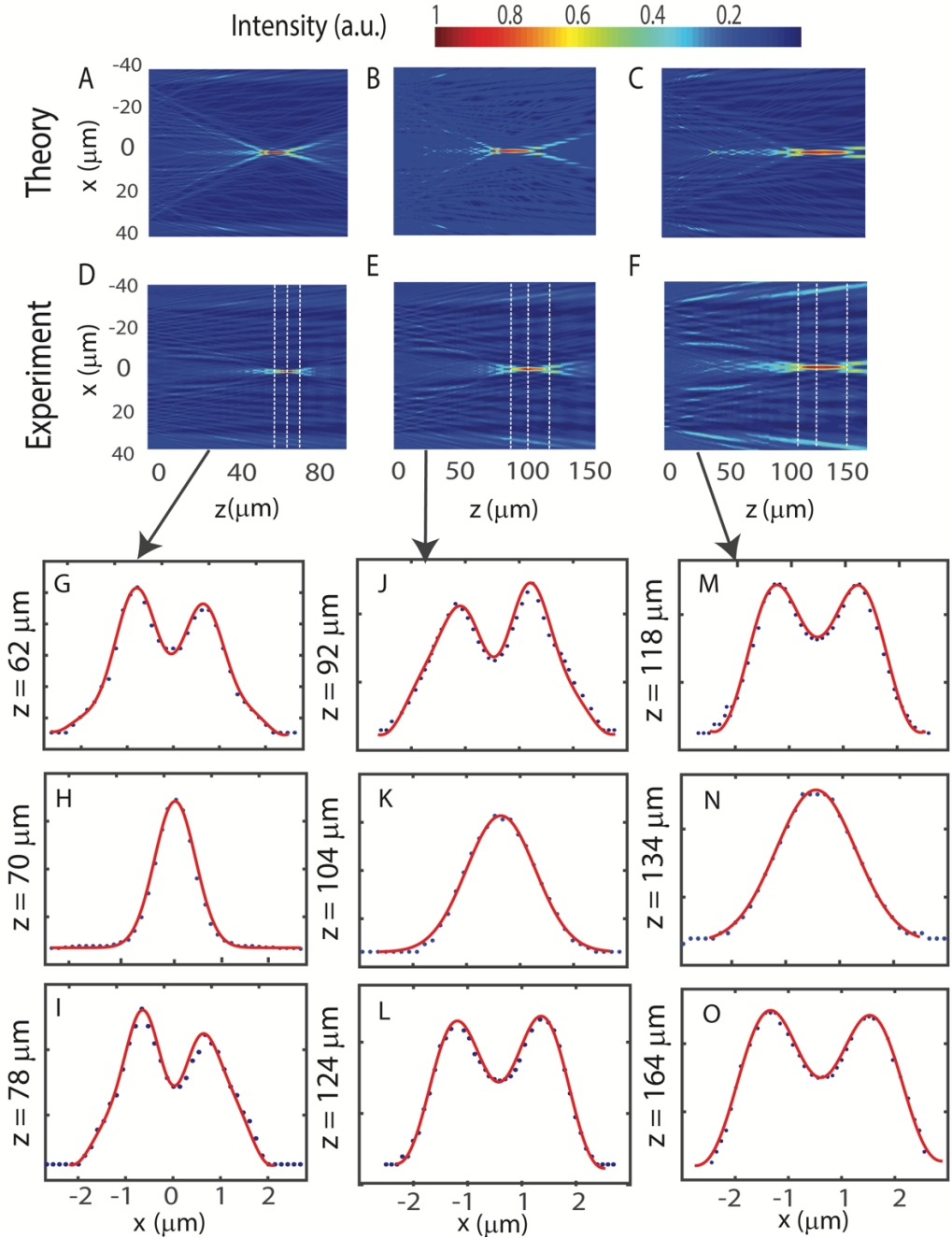

**Figure 2**: Simulated intensity along the optical axis of the designed EDOF metalenses with focal lengths of (A) 66.66 µm, (B) 100 µm, and (C) 133.33 µm. Experimentally measured field profile along optical axis for EDOF metalenses with focal lengths of (D) 66.66 µm, (E) 100 µm, and (F) 133.33 µm. (G-O) show cross-section of the beam size in different distance from the EDOF metalenses with a Gaussian and a double Gaussian fit.

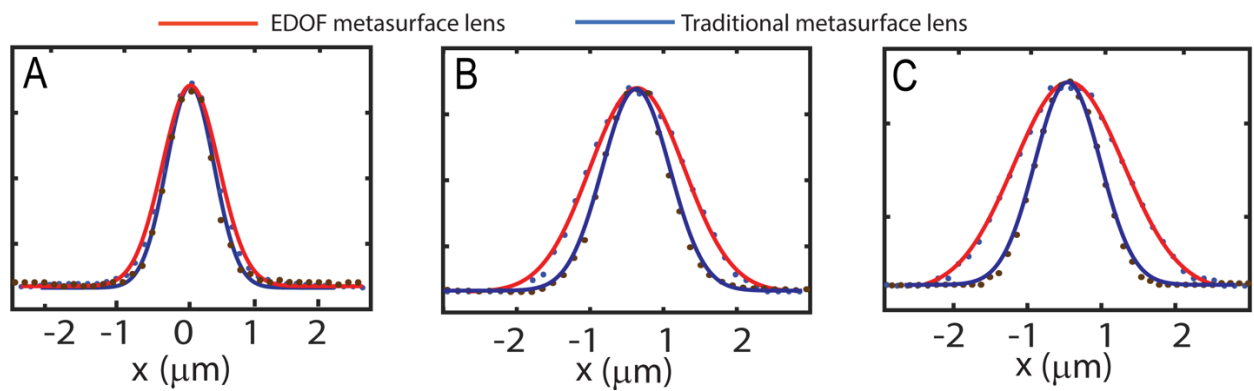

**Figure 3:** PSF of the traditional and EDOF metalenses for three different focal lengths: (A) 66.66 μm, (B) 100 μm, and (C) 133.33 μm.

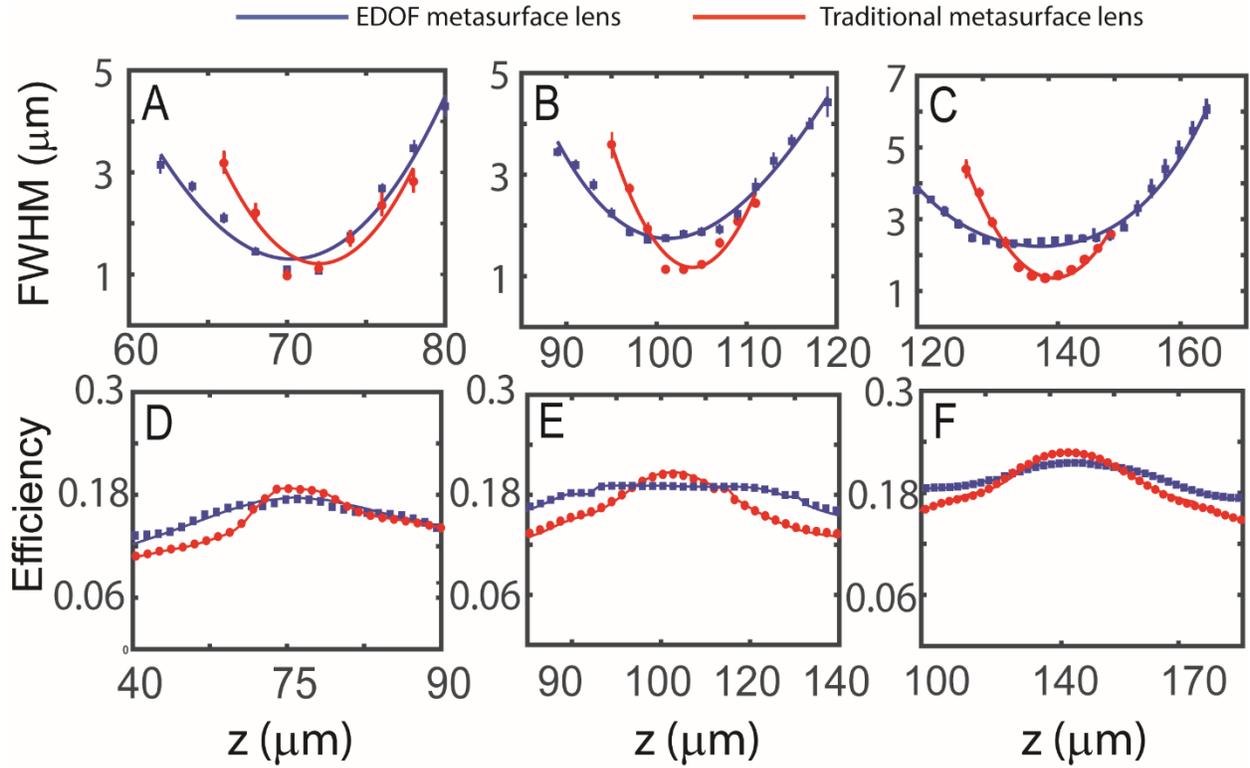

**Figure 4: Performance of fabricated ordinary and EDOF metalenses as a function of distance along the optical axis.** Measured FWHM of ordinary and EDOF metalenses as a function of distance corresponds to 66.66 μm (A), 100 μm (B), and 133.33 μm (C) metalenses. Measured focusing efficiency of 66.66 μm (D), 100 μm (E), and 133.33 μm (F) metalenses as a function of distance along the optical axis.

**Table 1: Comparison between the ordinary and EDOF metalenses for 66.66μm, 100μm, 133.33μm metalenses, respectively**

| Lens Properties | Ordinary Metalens | | EDOF Metalens | |
|---|---|---|---|---|
| | Simulation | Experiment | Simulation | Experiment |
| FWHM (μm) | 0.95 | 0.97 | 0.96 | 1.07 |
| | 1.1 | 1.13 | 1.3 | 1.7 |
| | 1.2 | 1.36 | 2.25 | 2.32 |
| Depth of Focus (μm) | 12($\pm$0.05) | 12($\pm$2) | 17($\pm$0.05) | 16($\pm$2) |
| | 14($\pm$0.05) | 16($\pm$2) | 30($\pm$0.05) | 30($\pm$2) |
| | 18($\pm$0.05) | 22($\pm$2) | 44($\pm$0.05) | 44($\pm$2) |
| Efficiency (%) | 19.1 | 18.6 | 16.8 | 16.34 |
| | 22.26 | 21 | 19.3 | 20.12 |
| | 24.18 | 24.65 | 22.21 | 23.48 |
| Focal Length (μm) | 67($\pm$0.05) | 70($\pm$2) | 64.2($\pm$0.05) | 70($\pm$2) |
| | 102($\pm$0.05) | 102($\pm$2) | 99.7($\pm$0.05) | 104($\pm$2) |
| | 136($\pm$0.05) | 138($\pm$2) | 129.8($\pm$0.05) | 134($\pm$2) |

# Supplement: Inverse designed metalenses with extended depth of focus

Elyas Bayati[1,*], Raphael Pestourie[2,3,*], Shane Colburn[1], Zin Lin[2], Steven G. Johnson[4], Arka Majumdar[1,4,+]

[1] Electrical and Computer Engineering, University of Washington, Seattle, WA 98189, USA
[2] Department of Mathematics, MIT, Cambridge, MA 02139, USA
[3] Harvard John A. Paulson School of Engineering and Applied Sciences, Harvard University, Cambridge, MA 02138, USA
[4] Department of Physics, University of Washington, Seattle, WA 98189, USA
[*] These authors contributed equally
[+] Corresponding Author: arka@uw.edu

4 pages, 2 figures S1-S2

## S1: Comparison with other potential EDOF lenses in terms of Focusing Performance

Other potential EDOF alternatives (e.g., axicons[1], log-aspheres[2], or cubic functions[3]) also exist for 1D lenses; however, different regions of these lenses focus at different depths, significantly limiting the focusing efficiency. We designed and simulated these EDOF cylindrical lenses via the forward design method[4]. The forward design method of these metalenses involves selecting the appropriate spatial phase profile (e.g., axicons, log-aspheres, or cubic functions) for the specific optical component, arranging the scatterers on a subwavelength lattice, and spatially varying their dimensions. To make a fair comparison with one of our inverse-designed lenses, other 1D EDOF lenses are designed to have central focal spot at 100 μm and a depth of focus of 30 μm. The field intensity profiles of these EDOF lenses are simulated using 2D finite-difference time-domain (FDTD) simulation with an axial sampling resolution of 50 nm (Figure S1). The focusing efficiencies of 1D log-asphere, axicon, and cubic lens were 12.16%, 8.6% and 14.47%, respectively. Even though, these 1D EDOF lenses can also provide an EDOF, their efficiency is significantly lower than that of our design, which is 20.12%. In this paper, the focusing efficiency is defined as the power within a circle with a radius of three times the FWHM at the focal plane to the total power incident upon the metalens Since FWHM, however, is not well defined for cubic lenses, we also calculated the focusing efficiency for this EDOF lens by taking the ratio of the intensity integrated within the aperture area of the device at the focal plane to that at the plane of metalens instead of a circle with a radius of three times the FWHM. Using this definition, we got 67% focusing efficiency for EDOF 1D cubic lens.

In addition of focusing efficiency, The FWHM of these 1D log-asphere, axicon and cubic lenses are 1.75μm, 1.25μm and 1.7μm, respectively which are calculated using their corresponding Gaussian fit at the cross sections of the focal planes. These FWHM are comparable to our inverse-designed metalens. Thus, our inverse design approach produces a metalens with comparable FWHM of other existing EDOF lenses, but maintains a higher focusing efficiency.

.

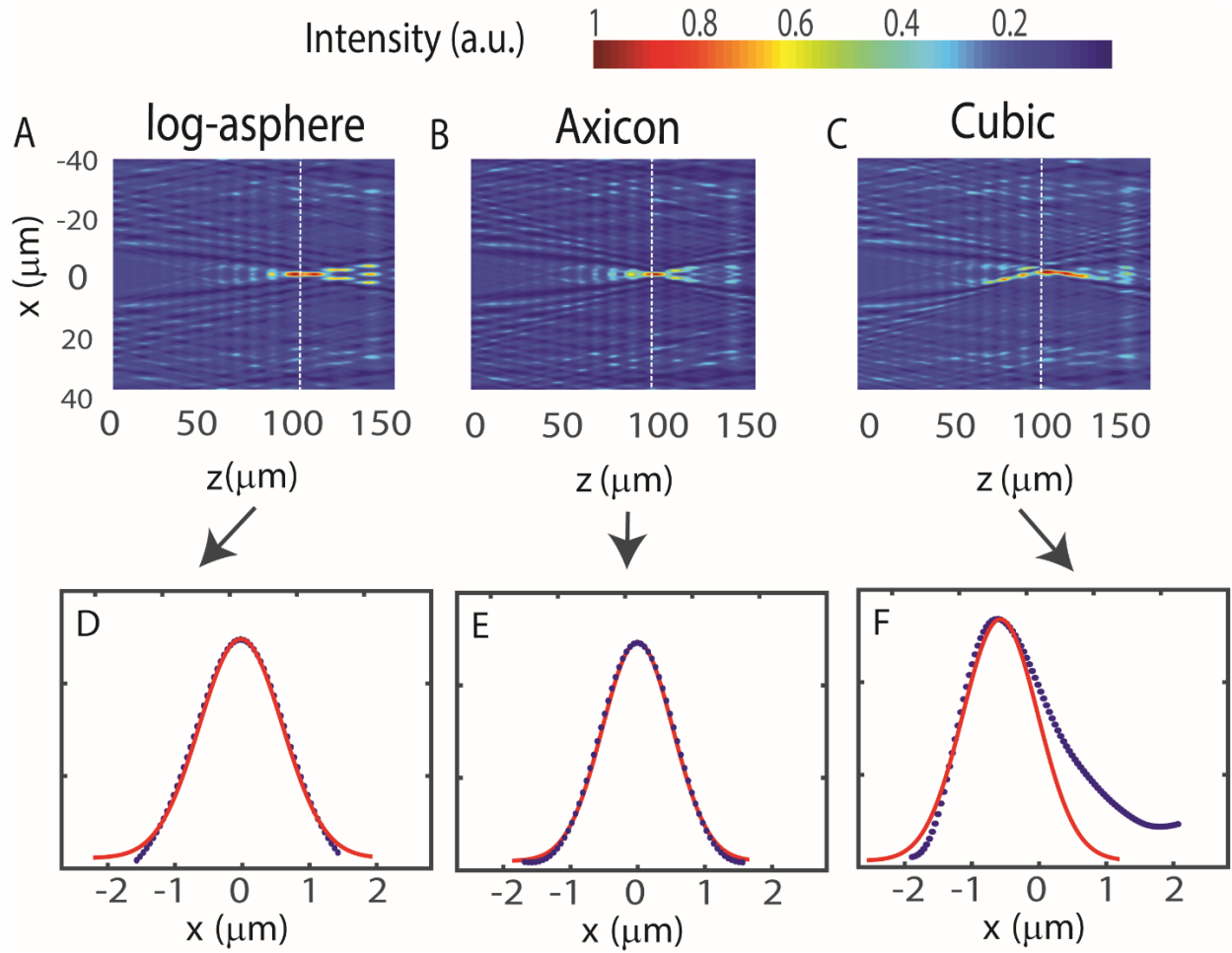

**Figure S1**: Simulated field intensity along the optical axis of alternative EDOF lenses with a focal length of 100 μm and depth of focus of 30 μm. We simulated intensity profiles for a log-asphere (A), axicon (B), and a cubic phase mask (C). (D-F) Cross sections at the focal plane of these EDOF lenses with a corresponding Gaussian fit. The focusing efficiency of 1D log-asphere, axicon and cubic lenses were calculated using these cross sections are 12.16%, 8.6% and 14.47%, respectively.

**S2: Simulated performance under oblique incidence**

In order to see how sensitive our inverse-designed EDOF lenses are to different incident angles, we also characterized the performance of the inverse-designed EDOF metalens under oblique incidence angles (5°, 10° and 15°) using FDTD simulation (Figure S2). While we clearly observe the effect of off-axis aberrations, the extended depth of focus remains the same for different angles. We emphasize, however, that as the inverse design figure of merit did not explicitly handle nonzero incident angles, we do not expect such aberrations to be mitigated in our design.

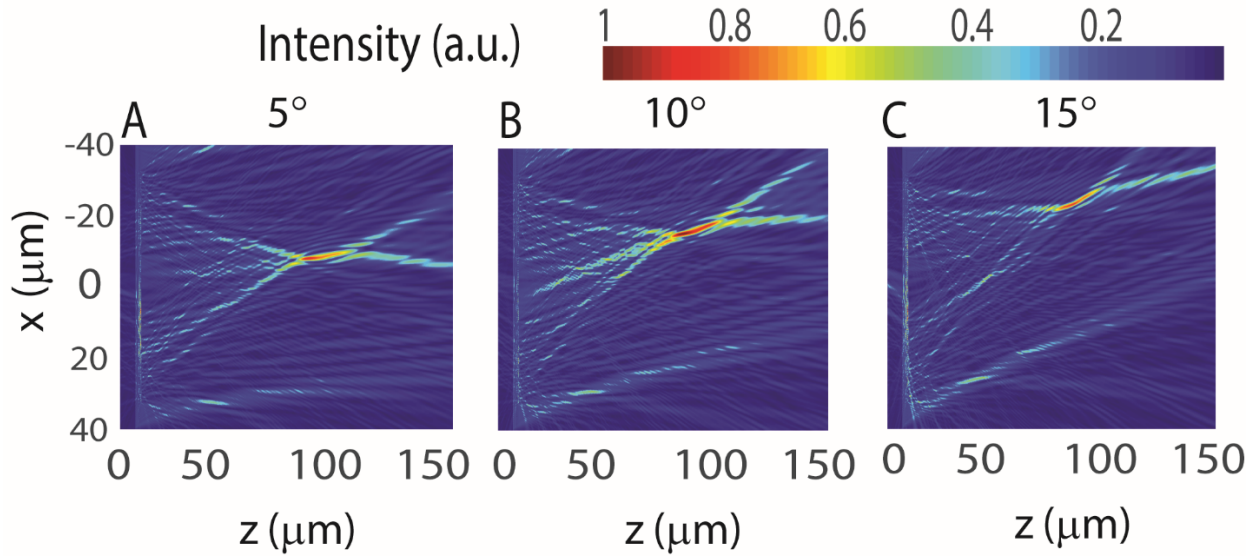

**Figure S2**: Simulated normalized field profile intensity of our inverse designed EDOF metalens with a focal length of 100 μm under oblique incidence. FDTD simulations were run for incidence angles of (A) 5 °, (B) 10 °, (C) 15 °. While we clearly observe the effect of off-axis aberrations in our simulation, there is a notable extension of depth of focus for different angles.